\documentclass[prc,12p,showpacs,showkeys]{revtex4-1}
\usepackage{graphicx}
\usepackage{dcolumn}
\usepackage{longtable}
\usepackage{amsmath}
\usepackage{amssymb}
\usepackage{natbib}
\usepackage{array}
\usepackage{multirow}
\usepackage{calligra}
\usepackage{float}
\usepackage{commath}

\begin{document}
\title{$\bf{A+2n}$ compound nuclei and the unitary limit in nuclear physics}
		\author{P E Georgoudis$^{1,2}$}
	
	\affiliation{$^1$Grand Accelerateur National d'Ions Lourds, CEA/DRF-CNRS/IN2P3, BP 55027, F-14076 Caen Cedex 5, France. $^2$ Department of Physics, National and Kapodistrian University of Athens, Zografou Campus, GR-15784 Athens, Greece.}
	\email{pageorgo@phys.uoa.gr}
		\begin{abstract}
		This contribution discusses a new perception of the structure of compound nuclei by introducing intermediate states of the Feshbach formalism of nuclear reactions in the Interacting Boson Model of nuclear structure. The stake is to explore the manifestation of the unitary limit in heavy, even-even nuclei. Interactions that govern Feshbach resonances of cold and dilute atomic gases suggest the formulation of an IBM-compound Hamiltonian for scattering two neutrons ($2n$) from a heavy, even-even target ($A$). The solutions of the corresponding coupled channel equations host a $2n$-IBM state resonance. It turns out that the unitary limit is measurable in a heavy $A+2n$ compound nucleus at low temperatures. That measurement is feasible through the fluctuations of the cross-sections that tune the $2n$-$A$ scattering length.
	\end{abstract}
	\maketitle
	\section{Introduction}
	Heavy, even-even nuclei serve as an inclusive testing ground for a remarkable pluralism of physical phenomena that range from Bardeen Cooper Schriefer behavior of nucleon pairs at valence shells to boson condensates for the emergence of the nuclear shapes as well as to quantum critical points in Shape/Phase transitions \cite{CJC}. The Interacting Boson Model provides a group theoretical framework that synthesizes a simple and integrated approach out of such a diversity of physical aspects in nuclear physics [2]. On the other hand, cold and dilute atomic gases manifest the so-called unitary limit  in the vicinity of Feshbach resonances \cite{Randeria}. The unitary limit serves as a theoretical benchmark that, like IBM, accompanies a similar but enriched diversity of physical aspects ranging from the BCS-Bose Einstein Condensation crossover to a quantum critical point. In parallel, the unitary limit refers to a strong coupling problem that is describable in terms of a Conformal Field Theory \cite{CFT}. The theoretical application of the unitary limit in nuclear physics was first realized in light nuclei \cite{light}. It emerged parallel with the development of Effective Field Theories \cite{EFTs}. This contribution briefly reviews the first exploration of the unitary limit in heavy even-even nuclei \cite{Georgoudis}. Apart from the phenomenological insights that this contribution focuses on, that exploration initiated the examination of conformal symmetry within the group theoretical framework of IBM. Conformal invariance arises at the critical point of a second-order phase transition \cite{Fisher} as well as in Quantum Chromo Dynamics at the limit of a large number of gluons \cite{Gross}. In that perspective, introducing the unitary limit in heavy even-even nuclei commences the investigation of algebraic relations between the symmetries of the IBM with the conformal symmetry of second-order critical points. In parallel, it explores algebraic relations with the limit of conformal invariance in a strong coupling problem (unitary limit) that is amenable to be incorporated afterward with QCD.
	
	\section{Lessons from the unitary limit in systems of cold atoms}
	Consider a scattering problem with a wavefunction that depends on the radial distance $r$ between the scattering particles and the phase shifts $\delta(k)$. In what follows, the discussion is restricted to $s$-wave scattering. One introduces the quantity of the scattering length $a$ through the behavior of the phase shifts at very low energies, that is $\delta(k)\sim-ka$ for $k \rightarrow 0$. For example, a scattering wavefunction of the form $\sin(kr+\delta(k))$ looks like $k(r-a)$ at very low energies. The scattering length $a$ shows the intercept of the wavefunction with respect to the horizontal axis, that is the radial distance. Apropos unitarity, one now focuses on the exhaustion of the unitarity bound in the cross-section. The parameter that controls the deviation from the unitarity bound is the generalized scattering length $a(k)$ as defined by Bethe \cite{Bethe} through the effective range ($r^{*}$) expansion
	\begin{equation}\label{e2}
		-k\cot\delta(k)=\frac{1}{a(k)}=\frac{1}{a}-\frac{1}{2}k^{2}r^{*}+ \cdots.
	\end{equation}
	The condition $1/a(k)=0$ on the generalized scattering length defines a resonance at the scattering amplitude \cite{Bethe}. At very low kinetic energies, $kr^{*}<<1$, that resonance condition on Eq (\ref{e2}) implies an infinite value for the scattering length, i.e. $a \rightarrow \infty$. The latter amounts to maximize the interaction strength between two particles, $g=4\pi a \hbar^{2}/m$ with $m$ the mass of each particle, i.e., it reflects a strong coupling limit. The unitary limit refers to the infinite value of the scattering length $a$ at low kinetic energy. A paradigm of its experimental observation has been achieved in the cold and dilute atomic gases \cite{Randeria}. 
	
	The open-closed channel crossing during an atom-atom collision is the underlying mechanism that realizes the unitary limit. The open channel reflects a scattering state of two cold atoms, while the closed channel is the bound state of a diatomic molecule formed by these same cold atoms. Channels' crossing means the coincidence in the energy of two different channels. In the Feshbach formalism for reactions, the open-closed channel crossing is achieved through the resonating energy of the open channel with the energy of an intermediate state of the closed channel. That resonating energy gives rise to a resonance that manifests the intermediate state of the closed channel. In atomic and molecular physics, these resonances are the celebrated Feshbach resonances  \cite{Timmermans}. However, they were initially introduced in compound nuclei \cite{Feshbach}.

	The intermediate state of the Feshbach formalism affects the wavefunction's scattering length $a$. In general, for low $k$, the element of the scattering matrix is expressed through the phase shifts as $S_{0}=e^{2i\delta(k)}=e^{-2ika}$. In the presence of an intermediate state of energy $E_{m}$ and width $\Gamma_{m}$, the quantity $S_{0}$ takes the form 
	\begin{equation}\label{e20}
		S'_{0}=e^{2ika}\left(1-i\frac{\Gamma_{m}}{E-E_{m}+i\Gamma_{m}/2} \right).
	\end{equation}
	From the perspective of nuclear physics \cite{Feshbach}, Eq. (\ref{e20}) generates the scattering matrix element  $S'_{0}=S_{0}S_{R}$, with 
	$S_{R}=1-i \Gamma_{m}/\left(E-E_{m}+i\Gamma_{m}/2\right)$ a fluctuating part that fluctuates rather rapidly with the energy by the resonating energies $E_{m}$ and widths $\Gamma_{m}$. Now, like in atomic and molecular physics \cite{Timmermans}, one identifies the effect of that fluctuating part in the emergence of an effective scattering length $a_{eff}=a+a'$, with
	\begin{equation}\label{e201}
		e^{2ika'}=1-i\frac{\Gamma_{m}}{E-E_{m}+i\frac{\Gamma_{m}}{2}}, \ a_{eff}=a+\frac{1}{2k}\tan^{-1}\left(\frac{\Gamma_{m}(E-E_{m})}{(E-E_{m})^{2}+\frac{\Gamma_{m}^{2}}{4}} \right).
	\end{equation}
	The effective scattering length goes to infinity at the resonating energies $E=E_{m}$ of the open channel with the intermediate states. Therefore, a Feshbach resonance maximizes the scattering length. Experimentally in ultracold atoms, an external magnetic field tunes the energy of the channels to achieve their crossing that generates the Feshbach resonance.
	
	That said, one devises now an approach to introduce the unitary limit in heavy, even-even nuclei. Their low-lying collective states are formed by nucleon pairs that behave as bosons and generate coherent states to produce the nuclear shapes \cite{IBM}. Therefore, one examines the formation of an IBM boson in analogy with a Feshbach resonance during the formation of a diatomic molecule out of two cold atoms. Such a resonance is introduced in \cite{Georgoudis} for scattering two slow neutrons ($2n$) with a heavy even-even nucleus of mass number $A$. In that case, the Feshbach formalism's intermediate states affect the scattering matrix's element by a fluctuating term  like in Eq. (\ref{e20}). Such fluctuations are experimentally testable in nuclear physics and represent the formation of the compound system. It is outlined below how the observation of a particular fluctuation centered around the two-neutron separation energy ($S_{2n}$) maximizes the neutron pair $(2n)$-IBM state scattering length. 
	\section{Algebraic correspondences between the IBM and systems of cold atoms}
	Werner and Castin \cite{Castin} introduced mappings between zero-energy states and trapped states widely used in cold atoms.  A trapped state is merely a quantized state of a harmonic oscillator. One opens the walls of the trap by reaching the zero-frequency limit $\omega=0$ in the harmonic oscillator and obtains the zero-energy state. Werner-Castin mappings were introduced in the solutions of the Schrodinger equation for N trapped cold atoms or particles in general. They preserve the unitary limit and are realized through the generators of the $SO(2,1)$ group in an isomorphic realization to the generators of the conformal group in one dimension - time.   An algebraic correspondence for those mappings is established with the simplest form of the $O(6)$ limit of the IBM \cite{Georgoudis}. In other words, the Werner-Castin mappings \cite{Castin} correspond to certain relations in the $O(6)$ limit of the IBM under the appropriate algebraic replacements. By this process, one introduces the one-dimensional conformal group in the IBM and writes down the wavefunction for the corresponding zero-energy state. This state contains a scaling exponent that arises out of the invariant quantity of dilatation (scale) transformations of the boson number radius $\rho$. That invariant quantity is the $O(6)$ quantum number $\sigma$. Table \ref{t1} summarizes the main replacements/relations of this correspondence.
	
	The practical result for nuclear physics emerges by noticing that Werner-Castin mappings apply to Feshbach resonances by generating the intermediate (trapped) state of the closed channel out of the channels' crossing (zero-energy state) and vice-versa. By applying those mappings to the intermediate states of the $A+2n$ compound nucleus, one obtains the corresponding zero-energy states that reveal the open-closed channel crossings. For instance, if the ground state of the target $A$ contains $N_{b}$ bosons at equilibrium, one applies the mapping to the intermediate state of the $A+2n$ compound nucleus with the target having $N_{b}+1$ bosons. The result is to {\it remove} one boson and take the zero-energy state in the form $\rho^{N_{b}+1}$. That state is the crossing of the open channel - consisting of the neutron pair $(2n)$ and the target nucleus of $N_{b}$ bosons - with the closed channel consisting of the $N_{b}+1$ bound state of the IBM formed by {\it adding} that $(2n)$ as one boson. 
	\section{A+2n compound nucleus at low temperatures like a cold and dilute atomic gas}
	\begin{table}
		\caption{Algebraic correspondences between the Schrodinger equation of $N=2$ trapped atoms with the IBM O(6) limit. $r_{1,2}$ is each atom's radial distance from the trap's center, and $l_{1,2}$ are their angular momenta. $L_{\pm}$, $L_{0}$ are the $SO(2,1)$ generators \cite{Georgoudis}.}\label{t1}
		\begin{center}
			\begin{tabular}{ c  c  c }
				\hline \\
				& $N=2$ hyperspherical & $O(6)$ IBM \\ \hline
				radial variable &  $R=\sqrt{r^{2}_{1}+r^{2}_{2}}$ & boson number radius $\rho$ \\ \hline
				dilatation eigenvalue & $\lambda= l_{1}+l_{2} $ & quantum number $\sigma$ \\ \hline
				energy & $\left(\lambda+2q+6/2\right)\hbar \omega$ & $\left(N_{b}+6/2\right)\hbar\omega$, $N_{b}=\sigma +2J$\\  \hline
				zero energy state & $\psi^{0}_{\lambda}=R^{\lambda+2q}$ & $\psi^{0}_{\sigma}=\rho^{N_{b}}$ \\\hline
				Werner-Castin mapping & $|F^{q}_{\lambda} \rangle = L^{q}_{+}e^{-R^{2}/2a^{2}_{ho}} |\psi^{0}_{\lambda} \rangle$ &  $|\Phi^{J}_{\sigma} \rangle = L^{J}_{+}e^{-\rho^{2}/2a^{2}_{ho}} |\psi^{0}_{\sigma} \rangle$ \\ \hline
			\end{tabular}
		\end{center}
	\end{table}
	The target nucleus' two neutron separation energy $S_{2n}$ determines the length scale. The boson number radius $\rho$ is measured in units of the harmonic oscillator length $a_{ho}=\hbar/\sqrt{M S_{2n}}$ where $M$ is the neutron mass. In that case, one writes down the radial distance between the $2n$ and the IBM state in the form of $R-\rho \equiv r$ and the corresponding wavenumber as $k_{r}$.
	
	Channel states are restricted to scalar angular momentum couplings of the form $\Psi(r,\rho)=\sum_{n}\Psi_{n}(r)\Phi_{n}(\rho)$ where $\Psi_{n}(r)$ is the $2n$ wavefunction, and $\Phi_{n}(\rho)$ is the IBM wavefunction. The scattering occurs at the cold limit, i.e., at a much lower kinetic energy of the $2n$ concerning the target's $S_{2n}$.  The open channel is the $n=0$, where the target contains $N_{b}$ bosons, while channel states with the target in higher boson numbers $N_{b}+1,2 \dots$ are those for $n>0$. These channel states form the Hilbert space of the IBM-compound Hamiltonian
	\begin{equation}\label{e8}
		H_{c}=H(r)+ H (\rho)+H(\rho,r).
	\end{equation} 
	$H(r)$ is the Hamiltonian for the relative kinetic energy of the $2n$ with respect to the target $A$ and the interaction between the two neutrons. $H(\rho)$ is the IBM Hamiltonian for the target, and $H(\rho,r)$ is the $2n$-IBM state interaction term.
	To investigate the unitary limit of $H_{c}$, the interaction terms are specified in analogy with the  interactions and the external magnetic field governing the vicinity of Feshbach resonances in cold and dilute atomic gases. There, when the formation of diatomic molecules reaches the unitary limit, the atom-atom unitary interaction induces a molecule-molecule unitary interaction \cite{Petrov}. In $H_{c}$, that means one introduces a unitary interaction for the incident neutrons themselves (atom-atom) plus a unitary interaction for the $2n$-IBM state coupling (molecule-molecule). The dilute character of the target's valence space concerning the short range of the strong interaction and the cold neutron pair rationalizes the analogy. Accordingly, a $2n$-IBM state scattering length $a_{r}$ is introduced through the corresponding effective range expansion as shown in Table \ref{t2}.
	
	One has the neutron-neutron scattering length $a$ and the pair-IBM state scattering length $a_{r}$ and the corresponding unitary interactions are 
	\begin{equation}\label{e9}
		\begin{split}
			&\frac{4\pi a \hbar^{2}}{M}\delta(r_{1}-r_{2}) \rightarrow \lim_{r_{1}\rightarrow r_{2}}\Psi_{0}(r)=\frac{C}{r_{1}-r_{2}}-\frac{1}{a}, \\
			&\frac{4\pi^{3}a_{r}\hbar^{2}}{M}\delta(r) \rightarrow \lim_{r\rightarrow 0}\Psi_{0}(r, \rho)=\Phi_{0}(\rho)\left(\frac{C}{r^{4}}-\frac{1}{a^{4}_{r}}\right).
		\end{split}
	\end{equation} 
	These boundary conditions apply to the $2n$ scattering wavefunction $\Psi_{0}(r)$, and to the channel wavefunction $\Psi_{0}(r, \rho)$. They replace the two unitary interactions, respectively. The full IBM-compound Hamiltonian now reads
	\begin{equation}\label{e10}
		H_{c}=-\frac{\hbar^{2}}{2M}\left(  \frac{1}{r^{5}} \frac{\partial}{\partial r}r^{5} \frac{\partial}{\partial r} - \frac{\lambda(\lambda+4)}{r^{2}} \right)+H(\rho)+s^{\dagger}+s.
	\end{equation}
	The $s^{\dagger}+s$ term changes target states by one $s$ boson. It is the analog of the magnetic field tuning to achieve the open-closed channel crossing. The effective range of the $2n$-IBM state interaction is the $r^{*}$ as seen in Table \ref{t2}. That range is detemrined experimentally by the width of the resonance that corresponds to the channels' crossing through the relation $\Gamma_{m}=\hbar^{2}k_{r}/Mr^{*}$.
	
	Intermediate states of the Feshbach formalism are {\it stationary} states of the $A+2n$ compound nucleus formed by the target plus two neutrons. They serve as resonance states of energy $\epsilon_{n}$ with respect to the total energy $E$ of the open channel. The coupled channels equations read
	\begin{equation}
		\begin{split}\label{e11}
			(E-H_{PP})P|\Psi \rangle=H_{PQ} Q |\Psi \rangle,\\
			(E-H_{QQ})Q|\Psi \rangle=H_{QP} P|\Psi \rangle.
		\end{split}
	\end{equation}
	The projection operators are the open channel $P=|\Phi_{0}(\rho)\rangle \langle \Phi_{0}(\rho)|$ and the set of closed channels $Q=\sum_{n>0}|\Phi_{n}(\rho)\rangle \langle \Phi_{n}(\rho)|$. Open-open $H_{PP}$ and closed-closed $H_{QQ}$ channel couplings are the unitary interactions and are included in the boundary conditions (\ref{e9}). We examine the coupling of the open channel ($n=0$) of $N_{b}$ bosons with the first closed channel ($n=1$) of $N_{b}+1$ bosons. The coupling $H_{PQ}$ now is $H_{10}=\langle N_{b}+1|s^{\dagger}+s|N_{b}\rangle=\sqrt{N_{b}+1}$ and the reverse $H_{QP}$ is the same $H_{01}=\langle N_{b}|s+s^{\dagger}|N_{b}+1\rangle=\sqrt{N_{b}+1}$.
	\section{Results}
	The energy scale is normalized to the energy of the target's ground state, i.e., to $N_{b}$ bosons.  The $2n$ $s$-wave ($\lambda=0$) open channel solutions are presented in sufficient detail in \cite{Georgoudis}. Table \ref{t2} summarizes the results for the $2n$ scattering compared to $1n$ scattering. The main difference is the cubic power in the solid angle factor of the cross-section and on the content of the wave numbers.
	
	In absence of a resonance with the intermediate state, the $s$-wave phase shifts of the $2n$-IBM state scattering  give the scattering matrix element $S_{0}=e^{2i\delta(k_{r})}=e^{-2ik_{r}a_{r}}$. Now, let us focus on the first closed channel where the energy of the target (IBM state) is denoted by the capital $E_{1}=(N_{b}+1+6/2)\hbar \omega$ and differs by $S_{2n}$ from the energy of the target in the open channel of $N_{b}$ bosons. The intermediate state $\Psi_{1}(r)$ of the neutron pair on that closed channel has, in general, an unknown energy denoted by $\epsilon_{1}$. Its Schrodinger equation is obtained by the second equation of (\ref{e11}) by setting $H_{10} = 0$. Then, the total energy $E$ is restricted to the energy $\epsilon_{1}$ of the intermediate state, and its equation reads
	\begin{equation}
		\begin{split}\label{e12}
			\left(T_{r}+E_{1}\right)\Psi_{1}(r)=\epsilon_{1}\Psi_{1}(r).
		\end{split}
	\end{equation}	
	This equation supports a zero-energy state under the condition $E_{1}=\epsilon_{1}$. That condition is satisfied when the energy of the intermediate state of the $2n$ in the compound $A+2n$ nucleus coincides with the bound IBM state of $N_{b}+1$ bosons. Unitarity manifests itself for that state by turning to the effect on the scattering matrix element. The new element of the scattering matrix reads $S'_{0}=S_{0}S_{R}$, with the fluctuating term $S_{R}=1-i \Gamma_{1}/\left(E-E_{1}+i\Gamma_{1}/2\right)$. Like in Eq. (\ref{e201}), that fluctuation generates an effective $2n$-IBM state scattering length $a_{reff}$ in the same sense with the emergence of the $a_{eff}$ in cold atoms. Therefore, at resonance, the condition $1/a_{r}(k_{r})=0$ applies and the $a_{reff}$ affects the $1/a_{r}$ part of the effective range expansion. That resonance is measurable through the fluctuation $S_{R}$, which generates the compound-elastic cross-section
	\begin{equation}\label{e13}
		\sigma_{ce}=\frac{(4\pi)^{3}}{k^{2}_{r}} \frac{\Gamma^{2}_{1}}{(E-E_{1})^{2}+(\Gamma_{1})^{2}/4}.
	\end{equation}
	The exhaustion of the unitarity bound occurs when the resonance's energy is the two neutron separation energy $E=E_{1}=S_{2n}$. The vicinity of the unitary limit is quantified by the width $\Gamma_{1}=b^{2}_{1} \left(\frac{4 M} {\hbar^{2}}\right) k_{r}$, with  $b^{2}_{1}=(N_{b}+1) \abs{ \int dr \Psi_{1}(r)  \Psi_{0}(r)}^{2}$ \cite{Georgoudis}. The latter depends on the neutron mass, the kinetic energy, and the boson number of the closed channel times a spectroscopic factor for the intermediate state. 
	\section{Conclusions}
	This work did not examine the conditions under which the incident $2n$ are captured as one boson by the target nucleus to form the $A+2n$ compound nucleus. Instead, this is the result of the investigation of unitarity in the $A+2n$ compound nucleus, i.e., that the capture of the slowly incident $2n$ as one boson maximizes the $2n$-IBM state scattering length $a_{r}$. The accompanied phenomenological insight is sufficiently important. Namely, the energies and the widths of the fluctuations of the cross-sections of $A+2n$ compound nuclei propose an experimental case study to examine the unitary limit in nuclear physics. The proposed IBM-compound Hamiltonian guarantees the $2n$ capture as one boson by the coupling term $s^{\dagger}+s$. How is the neutron-neutron scattering length related to the $a_{r}$? In cold atoms the atom-atom scattering length $a_{t}$ is related to the molecule-molecule scattering length $a_{m}$ by the empirical relation $a_{m}=0.6 a_{t}$ \cite{Petrov}. The maximization of the one induces the maximization of the other at resonance. In the same sense, one argues that the maximization of the $2n$-IBM state scattering length $a_{r}$ induces the maximization of the neutron-neutron scattering length. 
	\begin{table}
		\caption{The solutions of the open channel of the $2n$-$A$ scattering compared to those of $1n$-$A$ scattering. $r_{1,2}$ is the radial distance of each neutron with respect to the heavy $A$ core.}\label{t2}
		\begin{center}
			\begin{tabular}{ c  c  c }
			\\	\hline
				& One neutron & Two neutrons \\ \hline
				radial variable & $r_{1}$ & $r=R-\rho$, $R^{2}=r^{2}_{1}+r^{2}_{2}$ \\ \hline
				wavenumber & $k_{1}$ & $k_{r}$ \\ \hline
				Scattering wavefunction & $ \Psi_{0}(r_{1})=\frac{e^{-ik_{1}r_{1}}}{r_{1}}-  S_{0}\frac{e^{ik_{1}r_{1}}}{r_{1}}$ & $\Psi_{0}(r)=\frac{e^{-ik_{r}r}}{r^{5/2}}-  S_{0}\frac{e^{ik_{r}r}}{r^{5/2}}$ \\ \hline
				Effective Range Expansion & $\frac{1}{a_{1}(k_{1})}=\frac{1}{a_{1}}-\frac{1}{2}k^{2}_{1}r^{*}_{1} + \cdots $ & $\frac{1}{a_{r}(k_{r})}=\frac{1}{a_{r}}-\frac{1}{2}k^{2}_{r}r^{*} +\cdots$ \\ \hline
				cross - section &  $\sigma=\frac{4 \pi}{k^{2}+1/a_{1}^{2}(k)}$ & $\sigma=\frac{(4 \pi)^{3}}{k_{r}^{2}+1/a_{r}^{2}(k)}$ \\ \hline
			\end{tabular}
		\end{center}
	\end{table}

	\section{Acknowledgments}
	This research has received funding by the European Union's H2020 program, Marie Sklodowska Curie Actions - Individual Fellowships, Grant Agreement No 793900-GENESE 17.

\end{document}